\documentstyle[prl,aps,twocolumn]{revtex}
\large
\begin{document}
\draft
\twocolumn[\hsize\textwidth\columnwidth\hsize\csname @twocolumnfalse\endcsname

\title{
{\bf
Elementary excitations, exchange interaction and
spin-Peierls transition in CuGeO$_3$}\\
\vspace{0.5cm}
\large
Invited talk at the LT21, Prague 1996
(to be published in Czech. Journ. of Phys.)
}

\author{\underline{D. Khomskii}$^{a,b}$, W. Geertsma$^a$, and M.
Mostovoy$^{a,c}$}

\address{
$^a$ Groningen University, Nijenborgh 4, 9747 AG Groningen, the
Netherlands \\
$^b$ P. N. Lebedev Physical Institute, Leninski prosp.53,
Moscow, Russia \\
$^c$ G. I. Budker Institute of Nuclear Physics, 630090
Novosibirsk, Russia}

\date{\today}
\maketitle
\begin{abstract}
\widetext
\leftskip 54.8pt
\rightskip 54.8pt
The microscopic description of the spin-Peierls
transition in pure and doped CuGeO$_3$ is developed taking into
account realistic details of crystal structure.  It it shown that
the presence of side-groups (here Ge) strongly influences
superexchange along Cu--O--Cu path, making it antiferromagnetic.
Nearest-neighbour and next-nearest neighbour exchange constants
$J_{nn}$ and $J_{nnn}$ are calculated.  Si doping effectively segments
the CuO$_2$-chains leading to $J_{nn}({\rm Si})\simeq0$ or even
slightly ferromagnetic.  Strong sensitivity of the exchange
constants to Cu--O--Cu and (Cu--O--Cu)--Ge angles may be
responsible for the spin-Peierls transition itself
(``bond-bending mechanism'' of the transition).  The nature of
excitations in the isolated and coupled spin-Peierls chains is
studied and it is shown that topological excitations (solitons)
play crucial role.  Such solitons appear in particular in doped
systems (Cu$_{1-x}$Zn$_x$GeO$_3$, CuGe$_{1-x}$Si$_x$O$_3$) which
can explain the $T_{SP}(x)$ phase diagram.
\end{abstract}
\leftskip 54.8pt
\pacs{ }
]

\narrowtext

\section{Introduction}

The low-dimensional materials are known to be very susceptible to
various instabilities, such as formation of charge- or
spin-density waves.  Probably, the first one discussed is the
famous Peierls instability of one-dimensional metals: lattice
distortion with the new lattice period $2 \pi / Q$, where the
wave vector $Q = 2 k_F$ (if there is one electron per site, the
lattice dimerizes).  The lattice distortion opens a gap in
the electron spectrum at the Fermi surface, so that the energies
of all occupied electron states decrease, which drives
the transition.  It is also known that this instability survives
when we include the strong on-site Coulomb repulsion between
electrons (the, so called, Peierls-Hubbard model),
\vspace{-0.15cm}
\[
H = - \sum_{l,\sigma} \left( J_0 + \alpha
(u_l - u_{l+1}) \right) \left( c^{\dagger}_{l \sigma} c_{l+1
\sigma} +  h. c. \right)
\]
\vspace{-0.35cm}
\begin{equation}
+ U \!\sum_l c^{\dagger}_{l \uparrow} c_{l \uparrow}
c^{\dagger}_{l \downarrow} c_{l \downarrow} +
\sum_l \!\left( \frac{P_l^2}{2M} +
\frac{K}{2}(u_{l+1}\!-\!u_{l})^2 \right)
\end{equation}
Here the first term describes the dependence of the electron
hopping integral $t_{l,l+1}$ on the change of the distance $u_l -
u_{l+1}$ between the neighbouring ions and the last term is the
lattice energy (which after quantization becomes $\sum_q \omega_q
b^{\dagger}_q b_q$).  The dimensionless electron--lattice coupling
constant $\lambda = 4 \alpha^2 / (\pi t_0 K)$ determines the
magnitude of the lattice distortion and the energy gap.

When Coulomb repulsion is strong, $U \gg t_0$, and there is one
electron per site, we are in the limit of localized
electrons (Mott-Hubbard insulator) with effective
antiferromagnetic interaction (Spin-Peierls model),
\[
H_{eff} = \sum_l J_{l,l+1} \;{\bf S}_l \cdot {\bf S}_{l+1}
\]
\vspace{-0.3cm}
\begin{equation}
\label{Heff}
+ \sum_l \left( \frac{P_l^2}{2M} +
\frac{K}{2}(u_{l+1} - u_{l})^2 \right)\;\;,
\end{equation}
where the exchange constant $J_{l,l+1} = J_0 + \alpha^{\prime}
(u_{l} - u_{l+1})$, $J_0 = 4 t_0^2 / U$ and $\alpha^{\prime} = 8
t_0 \alpha / U$.  The dependence of $J_{l,l+1}$ on the distance
between neighbouring spins again leads to an instability, as
the result of which the spin chain dimerizes.  Physically it
corresponds to a formation of singlet dimers---the simplest
configuration in the valence bond picture.  This transition,
known as the spin-Peierls (SP) transition, was extensively
studied theoretically \cite{1,2,3} and was previously observed
experimentally in a number of quasi-one-dimensional organic
compounds, such as TTF-CuBDT $(T_{SP} = 12 K)$ or TTF- AuBDT
$(T_{SP} = 2.1 K)$ \cite{4}.

Recently the first inorganic spin-Peierls material CuGeO$_3$ was
discovered \cite{5}.  Since then much experimental data on
this material, both pure and doped (mostly by Zn and Si), was
obtained.  The spin chains in this compound are formed by CuO$_4$
plaquettes with common edge (See Fig.1).  They apparently play
the main role in the spin-Peierls transition with $T_{SP} = 14
K$.  However, as we will discuss below, the interchain
interaction is also very important here.  The interchain coupling
is provided both by Ge ions (along $b$-axis of the crystal) and
by the well separated apex oxygens (along $a$-axis; direction of
the chains coincides with the $c$-axis of a crystal).

Experimentally it is established that the strongest anomalies in
CuGeO$_3$, both in the normal phase and at the spin-Peierls
transition, occur not along the $c$-axis, but, rather
unexpectedly, along the other two directions, the strongest one
found along the $b$-axis \cite{6,7}.  For instance, the
anomalies in the thermal expansion coefficient and in
the magnetostriction along the $b$-axis are several times
stronger than along the direction of the chains \cite{7,8}.

Further interesting information is obtained in the studies of
doping dependence of various properties of CuGeO$_3$.  It was
shown that the substitution of Cu by nonmagnetic Zn, as well as
Ge by Si, leads initially to a rather strong reduction of
$T_{SP}$ \cite{9,10}, which according to some recent data
\cite{11} is flattened out at higher doping level.
Simultaneously, antiferromagnetic order develops at lower
temperatures, often coexisting with the SP distortion
\cite{11,12}.

The aim of the present investigation is to provide microscopic
picture of the properties of CuGeO$_3$ taking into account
realistic details of its structure.  We will address below
several important issues:
\begin{itemize}
\item
The detailed description of the exchange interaction (Why is
nearest neighbour exchange antiferromagnetic?);
\item
The sensitivity of the exchange constants to different types of
distortion and the resulting from that microscopic picture of
the spin-Peierls transition, which may be called bond-bending
model (Why the anomalies are strongest in the perpendicular
directions? Why is SP transition observed in CuGeO$_3$ and not in
many other known quasi-one-dimensional magnets?);
\item
The nature of elementary excitations in spin-Peierls systems in
general (Are they the ordinary singlet-triplet excitations?  How
are they influenced by the interchain interaction?);
\item
The mechanism by which doping affects the properties of SP
system (Why is the effect so strong?  Why does the
system develops antiferromagnetic order upon doping?).
\end{itemize}

These questions are raised by the experimental observations, and
we hope that their clarification will help both to elucidate some
general features of SP transitions and to build the detailed
picture of this transition in CuGeO$_3$.

\section{Exchange Interaction in CuGeO$_3$. Role of Side-Groups
in Superexchange}

The first question we would like to address is: why is the
nearest-neighbour Cu-Cu exchange interaction antiferromagnetic
at all?  The well-known Goodenough-Kanamori-Anderson rules state,
in particular, that the $90^{\circ}$-exchange between two
half-filled orbitals is ferromagnetic.  In CuGeO$_3$ the Cu-O-Cu
angle $\theta$ in the superexchange path is $98^{\circ}$, which
is rather close to $90^{\circ}$.  Usually, the antiferromagnetic
character of the exchange is attributed to this small $8^{\circ}$
difference.  Our calculation [13], however, shows that it is not
enough: even in this realistic geometry the exchange constant for the
nearest neighbour (nn) spins for realistic values of parameters
(such as copper-oxygen overlap, magnitude of the Coulomb
interaction on copper and oxygen, Hund's rule intraatomic
exchange, etc) is still slightly ferromagnetic, $J^{c}_{nn} = -
0.6$meV.

To explain the observed values of $J^{c}_{nn}$ the idea was put
forward in \cite{13} that the antiferromagnetic coupling may be
enhanced by the, initially ignored, side-groups effect (here Ge).
As it is clear from Fig.1, there is a Ge ion attached to each
oxygen in CuO$_2$ chain.  The Coulomb interaction with Ge$^{4+}$
and the hybridization of $2p_y$ orbital of oxygen with Ge (see
Fig.2) destroys the equivalence of $p_x$-orbitals (shaded) and
$p_y$-orbitals (empty) of oxygen, which for $90^{\circ}$-exchange
was responsible for the cancellation of the corresponding
antiferromagnetic contributions to superexchange.  As a result, the
exchange with partial delocalization into Ge may become
antiferromagnetic even for $90^\circ$ geometry (for a detailed
discussion see \cite{13}).  The calculation gives a reasonable
value for the nearest-neighbour exchange interaction: $J^{c}_{nn} =
11.6$meV (the experimental value is $9 \div 15$meV, depending on the
procedure of extraction of $J^{c}_{nn}$ from the experimental
data).

We also calculated other exchange constants using a similar
approach.  For the interchain interaction along $b$ and $a$ axes
we obtained $J^{b}_{nn} = 0.7$meV and $J^{a}_{nn} = -3 \cdot
10^{-4}$meV, so that $J^{b}_{nn} / J^{c}_{nn} \approx 0.06$, and
$J^{a}_{nn} / J^{c}_{nn} \approx - 3 \cdot 10^{-5}$.  The
experimental values are: $J^{b}_{nn} / J^{c}_{nn} \approx 0.1$ ,
$J^{a}_{nn} / J^{c}_{nn} \approx - 0.01$.  Thus our theoretical
results are not so far from the experiment for the interchain
interaction in the $b$-direction and too small for $a$-axis.  We
note, however, that the ferromagnetic exchange in $a$-direction
is in any event very weak and a small variation of parameters
used in our calculation can easily changes this value quite
significantly.

More interesting is the situation with the next-nearest-neighbour
(nnn) interaction in the chain direction $J^{c}_{nnn}$.  As is
clear from Figs.1 and 2, there is a relatively large overlap of
the $p_x$ orbitals on neighbouring plaquettes, which leads to a
rather strong antiferromagnetic nnn coupling.  Our calculation
gives $\gamma = J^{c}_{nnn} / J^c_{nn} \approx 0.23 \div 0.3$.
From the fit of $\chi(T)$ curve Castilla {\em et al} \cite{15}
obtained the value $\gamma \approx 0.25$.  Note also that a
sufficiently strong nnn interaction may lead to a singlet
formation and creation of a spin gap even without the
spin-lattice interaction.  Such a state is an exact ground state
at the Majumdar-Ghosh point $\gamma = 0.5$ [16].  The critical
value for appearance of a spin gap is $\gamma \approx 0.25$
\cite{15}.  Thus, from both the fit to experimental data and our
calculations it appears that CuGeO$_3$ is rather close to the
critical point, so that one can conclude that both the
frustrating nnn interaction and the spin-lattice interaction
combine to explain the observed properties of CuGeO$_3$ (see also
\cite{8}).

Anticipating the discussion below, we consider here the
modification of the exchange constants caused by doping.  In
particular, we calculated the change of $J^{c}_{nn}$ when Ge
ion attached to a bridging oxygen is substituted by Si.  As
Si is smaller than Ge, one can expect two consequences.
First, it will pull closer the nearby chain oxygen, somewhat
reducing the corresponding Cu-O-Cu angle $\theta$.  The second
effect is the reduced hybridization of $2p_y$ orbital of this
oxygen with Si.  According to the above considerations (see also
\cite{13}) both these factors would diminish the
antiferromagnetic nn exchange.  Our calculation shows \cite{17}
that for realistic values of parameters the resulting exchange
interaction becomes either very small or even weakly
ferromagnetic, $J^{c}_{nn} = 0\pm1$meV.  Thus Si doping
effectively interrupts the chains similar the effect of
substituting Cu by Zn.  This result will be used
later in section 5.

\section{Bond-Bending Model of the Spin-Peierls Transition in
CuGeO$_3$}

We return to the discussion of the exchange interaction and its
dependence on the details of crystal structure of CuGeO$_3$. As
follows from the previous section, the largest exchange constant
$J^{c}_{nn}$ is very sensitive to both Cu-O-Cu angle $\theta$
and to the side group (here Ge). As to the second factor, one
has to take into account that, contrary to a simple model of
CuGeO$_3$
shown in Fig.2, in the real crystal structure Ge ion lies not
exactly in the plane of CuO$_2$ chain: the angle $\alpha$ between
Ge and this plane is $\sim 160^{\circ}$.
The actual crystal structure may be schematically depicted
as in Fig.3, where the dashed lines represent CuO$_2$-chains.

One can easily understand that $J^{c}_{nn}$ is also very
sensitive to a Ge-CuO$_2$ angle $\alpha$.  The influence of Ge,
which according to the above consideration gives an
antiferromagnetic tendency, is the largest when $\alpha =
180^{\circ}$: just in this case the inequivalence of $2p_x$ and
$2p_y$ orbitals shown in Fig.2, which is crucial for this effect,
becomes the strongest.  On the other hand, if, for instance,
$\alpha = 90^{\circ}$ ({\em i.e.} if Ge would sit exactly above
the oxygen) its interaction with $2p_x$ and $2p_y$ orbitals would
be the same and the whole effect of Ge on $J^{c}_{nn}$ would
disappear.  Thus bending GeO-bonds with respect to CuO$_2$-plane
would change $J^{c}_{nn}$ (it becomes smaller when $\alpha$
decreases).

These simple considerations immediately allows one to understand
many, at first glance, strange properties of CuGeO$_3$ mentioned
in the introduction \cite{19}.  Thus, {\em e.g.} the compression
of CuGeO$_3$ along the $b$-direction would occur predominantly by
way of decreasing of Ge-(CuO$_2$) angle $\alpha$, while the
tethrahedral O-Ge-O angle $\phi$ is known to be quite rigid.
Such a ``hinge'' or ``scharnier'' model explains why the main
lattice anomalies are observed along the $b$-axis \cite{7} and
why the longitudinal mode parallel to $b$ is especially soft
\cite{6}.  Within this model one can also naturally explain (even
quantitatively) the fact that the magnetostriction is also
strongest in the b-direction \cite{8}.  If we assume
that the main changes in the lattice parameters occur only due to
bond bending ({\em i.e.} due to the change of angles, while bond
lengths remain fixed), we obtain the following
result for the uniaxial pressure dependence of $J \equiv
J^{c}_{nn}$ \cite{19}: $\delta J/\delta P_b =
-1.5$meV/GPa, which is close to the experimental result
$\delta J/\delta P_b = -1.7$meV/GPa \cite{8}.
We can also explain reasonably well the change of the
exchange coupling for other directions.

This picture can be also used to explain the spin-Peierls
transition itself. What occurs below $T_{SP}$, is mostly the
change of bond angles (``bond-bending''), which alternates along
the chains.  Experimentally it was found \cite{20} that the
dimerization is accompanied by the alternation of Cu-O-Cu angles
$\theta$.  In our model $J$ is also sensitive to Ge-CuO$_2)$
angle $\alpha$ and we speculated in Ref.14 that this angle, most
probably, also alternates in the spin-Peierls phase.  Recently
this alternation was observed \cite{18}.

Consequently we have a rather coherent picture of the
microscopic changes in CuGeO$_3$, both above and below $T_{SP}$:
in the first approximation we may describe the main lattice
changes as occurring mostly due to the change of the ``soft'' bond
angles.  The strongest effects for $T>T_{SP}$ are then expected along
the $b$-axis, which is consistent with the experiment. The same
bond-bending distortions seem also to be responsible for the
spin-Peierls transition itself, the difference with the normal phase
being the alternation of the corresponding angles in the
$c$-direction.

The bond-bending model allows one to explain another puzzle
related to spin-Peierls transitions (discussed already in
\cite{2}): up to now such transitions have been observed only in a
few of the many known quasi-one-dimensional antiferromagnets.
There might be several reasons for that.
The first one is that the
spin-Peierls phase in CuGeO$_3$ is, at least partially,
stabilized by the frustrating next-nearest neighbour interaction
$J^{c}_{nnn}$.  The other factor is that the spin-Peierls
instability is greatly enhanced when the corresponding phonon
mode is soft enough \cite{2}.  One can see it {\em e.g.} from the
expression for $T_{SP}$ \cite{3},
\[
T_{SP} = 0.8 \lambda^{\prime} J\;\;.
\]
The spin-phonon coupling constant is
\[
\lambda^{\prime} = \frac{{\alpha^{\prime}}^2}{JK} =
\frac{{\alpha^{\prime}}^2}{J M \omega_0^2}\;\;,
\]
where $\omega_0 = \sqrt{K / M}$ is the typical phonon
frequency.

There is, usually, a competition between the $3d$ magnetic
ordering and the spin-Peierls phase.  Apparently, in most
quasi-one-dimensional compounds the $3d$ magnetic ordering wins,
and for the spin-Peierls transition to be realized a strong
spin-lattice coupling,{\em i.e.} rather soft phonons, is
necessary.  Such soft phonon modes are known to exist in the
organic spin-Peierls compounds \cite{2}.  In CuGeO$_3$ it can be
rather soft bond-bending phonons, especially the ones parallel to
the $b$-axis, which help to stabilize the spin-Peierls phase
relative to the $3d$ antiferromagnetic one.  Nevertheless, a
relatively small doping is sufficient to make the
antiferromagnetic state more favourable, although some other
factors are also very important here, as will become clear in
the next section.

\section{Solitons and Strings in Spin-Peierls Systems}

Let us turn now to the second group of problems related to SP
systems, namely, the nature of elementary excitations.  In the
simple picture mentioned in the introduction (and valid in the
strong coupling limit) the SP state consists of isolated dimers.
For the rigid dimerized lattice an excited state is a triplet
localized on one of the dimers and separated from the ground
state by an energy gap $J$.  The interaction between the
neighbouring dimers gives a certain dispersion to this
excitation, transforming it into an object similar to a usual
magnon.

If, however, the lattice is allowed to adjust to a spin flip, the
localized triplet decays into a pair of
topological excitations.  Such excitations (solitons or kinks)
are known to be the lowest energy excitations in electronic
Peierls insulators \cite{21}.  The same is also true for
spin-Peierls systems.  Indeed, there exist two degenerate ground
states in a SP chain: one with singlets formed on sites
\ldots(12)(34)(56)\ldots, and another of the type
\ldots(23)(45)(67)\ldots.  One can characterize them by the phase
of the order parameter $\phi_n$, so that $\phi_n = 0$ in the
first state and $\phi_n = \pi$ in the second.
The soliton is an excited state,
in which the order parameter interpolates from $0$ to $\pi$ or
vice versa.  In the strong coupling limit such a state looks like
\ldots(12)(34)$\uparrow$(67)(89)\ldots, see Fig.~4.  Actually, the soliton
has a finite width, which (as the correlation length in the BCS
theory) has a form,
\begin{equation}
\xi_0 \left(= \frac{\hbar
v_F}{\Delta_0}\right) \sim \frac{ J}{ \Delta} a \sim
\frac{J}{E_s} a \;\;.
\end{equation}
Here the Fermi velocity
$v_F \sim J a / \hbar$ is the velocity of the spinless
Jordan-Wigner fermions, in terms of which the Hamiltonian (2) has
a form similar to the Hamiltonian of the electronic Peierls
system, $2 \Delta$ is the energy gap and $a$ is the lattice
constant, which below we will put equal to $1$.  The excitation
energy of the SP soliton, $E_s$, can easily be determined for
the $XY$-model, {\em i.e.} if one ignores ${\bf S}^{z}_{l} \cdot
{\bf S}^z_{l+1}$ term in the Hamiltonian (\ref{Heff}).  Then the
spin-Peierls Hamiltonian (\ref{Heff}) after the Jordan-Wigner
transformation acquires a form of the Su-Schrieffer-Heeger
Hamiltonian \cite{21} for electronic Peierls materials, in which
case $E_s = \frac{2}{\pi} \Delta$ \cite{21}.  The omitted term
renormalizes the soliton energy, as well as the mean-field energy
gap $2 \Delta$, but these numerical changes do not play an
important role.  One should also note that the kinks are mobile
excitations with the dispersion $\sim E_s$.  In CuGeO$_3$ $\xi_0$
is estimated to be of the order of $8$ lattice spacings.

From Fig.4 it is clear that a soliton in SP system corresponds to
one unpaired spin.  Thus, these elementary excitations have
$\frac{1}{2}$ rather then $1$ as the singlet-triplet excitations.
Of course, for fixed boundary conditions the solitons (excited
{\em e.g.} thermally or optically) always appear in pairs.

So far we considered the excitations in an isolated SP chain.
Now we want to include the effects of the interchain interaction.
Due to this interaction (mediated, for instance, by
three-dimensional phonons) SP distortions of neighbouring chains
would prefer to be phase coherent, {\em e.g.} in phase.  When a
kink-antikink pair of size $r$ is created in one of the chains,
the phase of the distortion between the kink and antikink is
opposite to the initial one as well as to those on neighbouring
chains, which would cost an energy $E(r) = Z \sigma r$, where
$\sigma$ is the effective interaction between the Peierls phases
on different chains per one link and $Z$ is the number of
neighbouring chains (See Fig.5a).  Therefore, in the presence of
the interchain interaction the soliton-antisoliton pair forms a
string and $Z \sigma$ may be called the string tension.  The
linear potential of the string confines the soliton motion, {\em
i.e.} kink and antikink can not go far from each other in an
ordered phase.

We can use this picture to estimate the value of the temperature
of the $3d$ SP transition.  The concentration of thermally
excited kinks in an isolated chain is $n = \exp (- E_s / T)$ and
the average distance between them is ${\bar d(T)} = n^{-1} = \exp
(E_s / T)$.  At the same time, the average distance between the
kinks connected by string is ${\bar l}(T) = T / (Z \sigma)$.  The
three-dimensional phase transition (ordering of phases of the
lattice distortions of different chains) occurs when ${\bar l}(T)
\sim {\bar d(T)}$, {\em i.e.}
\begin{equation}
\frac{T_{SP}}{Z \sigma}
\sim \exp \left( \frac{E_s}{T_{SP}} \right) \;\;,
\end{equation}
or
\begin{equation}
\label{TSP}
T_{SP} \sim \frac{E_s}{\ln \frac{E_s}{Z \sigma}}
\sim \frac{\lambda^{\prime} J}
{\ln \left(
\frac{\lambda^{\prime} J}{Z \sigma} \right)}
\end{equation} where
we use the relation $E_s \sim \Delta \sim \lambda^{\prime} J$
\cite{3}.  In this picture at $T < T_{SP}$ the phases of SP
distortions of different chains are correlated and all solitons
are paired.  At $T > T_{SP}$ local SP distortions
still exist in each chain, but there is no long range order.
Therefore, the SP transition in this picture is of a
``deconfinement'' type, which is somewhat similar to the
Kosterlitz-Thouless transition in $2d$-systems.

The approach described above is valid when the value of the
interchain interaction $\sigma$ is much smaller than $J$.  Using
Eq.(\ref{TSP}) with $J = 100$K and $\lambda^{\prime} \sim 0.2$
\cite{5} we get $\sigma \sim 0.04J$, which in view of the
logarithmic dependence of $T_{SP}$ on $\sigma$ in (\ref{TSP}) is
just enough for applicability of the results presented above
(these are, of course, only an order of magnitude estimates).

\section{Solitons in Doped Systems}

As we have seen above, Zn and Si, the two most studied dopands of
CuGeO$_3$, lead to an effective interruption of spin chains into
segments of finite length.  The segments with even number of Cu
ions can have a perfect SP ordering, while the odd segments
behave differently: one spin $\frac{1}{2}$ remains unpaired,
which means that the ground state of an odd segment contains a
soliton (similarly to what happens in the electronic Peierls
materials \cite{SU}).  One can show that the soliton is repeled
by ends of the segment, and in an isolated odd segment the
situation would look like in Fig.4: the soliton carrying spin
$\frac{1}{2}$ would prefer to stay in the middle of a segment.
This conclusion is in contrast with the usual assumption that the
magnetic moments induced by doping are localized near the
impurities.

The situation, however, changes when we take into account the
interchain interaction.  As we have seen in the previous section,
moving a soliton along a chain costs an energy which grows
lineary with the distance.  As is illustrated in Fig.5a, this
provides a force pulling the soliton back to the impurity.  Thus
the soliton moves in a potential shown in Fig.5b: it repels from
the impurity with the potential $V_{imp}(r) \sim J \exp (- r /
\xi_0)$, while the interchain interaction gives the potential
$V_{conf}(r) \sim Z \sigma r$, providing the restoring force.  As
a result, the soliton is located at a distance $\sim \xi_0$ from
impurity, so that, in a sense, we return to the traditional
picture.  One should keep in mind, however, that for a weak
interchain interaction the total potential $V_{imp} + V_{conf}$
is rather shallow and at finite temperature the soliton can go
rather far from impurity.  It seems that it should be possible to
check this picture experimentally, {\em e.g.} by detailed NMR
study of doped SP compounds (cf.  the results of M.  Chiba {\em
et al}, this conference).

\section{Phase Diagram of Doped Spin-Peierls Systems}

One can use this picture to describe qualitatively
the dependence of the phase transition temperature $T_{SP}$ on
the concentration of dopands $x$.  Similar to the treatment given
in section 4, we compare an average distance between the kink and
the nearest end of the segment ${\bar l}(T) \sim T / (Z \sigma)$
with the average length of the segment ${\bar d} \sim 1 / x$.
This gives,
\begin{equation}
\label{T(x)}
T_{SP}(x) \sim \frac{Z \sigma}{x} \;\;.
\end{equation}

This result has two limitations.  At large $x$, when an average
length of the segment becomes of the order of soliton size, $1 /
x \sim \xi_0$, there will be no ordering even at $T = 0$.  Thus
$x \sim \xi_0^{-1}$ is an absolute limit beyond which the $3d$
ordering disappears.  Using our estimate $\xi_0 \sim 8$,
such $x_{max} \sim 15 \%$.  On the other hand, the result
(\ref{T(x)}) is also not valid at very small $x$.  When an
average size of segment ${\bar d(x)} \sim 1 / x$ becomes
sufficiently large, the thermally induced solitons become as
important as the solitons induced by disorder.  In this case the total
concentration of solitons is
\begin{equation}
n_{tot} = n_{imp} + n_{therm} = x + e^{- \frac{E_s}{T}}\;\;,
\end{equation}
and one should compare ${\bar l}(T)$ with
$n_{tot}^{-1}$. For $x = 0$ we return to
Eq.(\ref{TSP}), while for small $x$ we get,
\begin{equation}
\label{small}
T_{SP}(x) = T_{SP}(0) (1 - \alpha x)\;\;,
\end{equation}
where the coefficient $\alpha$ is
\begin{equation}
\alpha \sim \frac{E_s}
{Z \sigma \ln \left( \frac{E_s}{Z \sigma} \right)}.
\end{equation}

One can verify these estimates more rigorously by mapping the
spin-Peierls system onto an effective Ising model.  Let us
associate the classical Ising variable $\tau = \pm 1$ with the
two possible types of SP ordering (phases $0$ and $\pi$), so that
the phase $0$ (left domain in Fig.4) corresponds to $\tau = + 1$,
while the phase $\pi$ (right domain in the same figure)
corresponds to $\tau = - 1$.  In this language a soliton is
a domain wall in Ising variables.  Since it costs an energy
$E_s$ to create a soliton, the Hamiltonian of the intrachain
interaction in the effective Ising model can be written as,
\begin{equation}
H_{intra} = - \frac{E_s}{2} \sum_{n,\alpha}
\left( \tau_{n,\alpha} \tau_{n+1,\alpha} - 1 \right) \;\;,
\end{equation}
(here $\alpha$ is the chain index and $n$ is the site number in
chain). Similarly, an interchain interaction in terms of Ising
variables has a form,
\begin{equation}
H_{inter} = - \frac{\sigma}{2} \sum_{n,\alpha} \tau_{n,\alpha}
\sum_{\delta} \tau_{n,\alpha+\delta}\;\;,
\end{equation}
where the summation over $\delta$ goes over neighbouring chains.
One can also introduce impurities in this effective Ising model.
The detailed treatment of this model will be given in a
separate publication \cite{22}.  Here we limit ourselves by
presenting in Fig.6 the results of the numerical solution of the
equation for the transition temperature for several values of the
interchain interaction $\sigma$. The value of $J$ was adjusted to
make the transition temperature equal to $14$K for each value of
$\sigma$. The values of $\sigma$ (from the top curve to the bottom
one) are (in K) 3; 2; 1; 0.5; 0.1. We see that the behaviour of
$T_{SP}(x)$ agrees with the (\ref{T(x)}) at large $x$ and
(\ref{small}) at small $x$ and for the values of the parameter
$\sigma$ not much different from the estimates made in section 4 one
can obtain a reasonable form of the phase diagram for CuGeO$_3$.
(One should also take into account that each Ge is coupled to two
chains, so that its substitution by Si introduces two
interruptions in exchange interaction, whereas Zn interrupts only
one chain.)

As follows from our picture, each soliton introduced by doping
carries an uncompensated spin $\frac{1}{2}$.  One can easily show
that in the vicinity of the domain wall where the SP order
parameter is small there exist antiferromagnetic spin
correlations (see Fig.7).  Both these correlations and the SP
distortion change on a length scale $\xi_0$.
Antiferromagnetic correlations on neighbouring kinks may overlap,
which could, in principle, lead to the long-range antiferromagnetic
ordering.  Thus it is possible to obtain a regime in which
the SP and antiferromagnetic orderings coexist.  To study this
question in detail one must also take into account also the
interchain exchange interaction.  This question is now under
investigation \cite{23}.

\section{Concluding Remarks}

To summarize we have a rather coherent picture of the main
properties of the SP system CuGeO$_3$.  The treatment given in
the first part of this paper allows one to explain many of the
features of this compound, which, at first glance, look rather
puzzling, such as the strong anomalies observed in the direction
perpendicular to chains rather than parallel to them.
Furthermore we showed how the local geometry and the side-groups
(Ge, Si) lead to a rather detailed microscopic
picture of the distortions in CuGeO$_3$ both above and below
$T_{SP}$.  These results are largely specific for this particular
compound, although some of the conclusions ({\em e.g.} the role
of side groups in superexchange and the importance of the soft
bending modes) are of a more general nature.

The results of the second part of the paper, though inspired by
the experiments on CuGeO$_3$, have a general character, {\em
e.g.} the conclusions about the domain wall structure of the
elementary excitations, confinement of solitons caused by the
interchain interaction, disorder-induced solitons \cite{MFK},
etc.  At the same time, this general treatment provides a
reasonable explanation of the suppression of $T_{SP}$ by doping
and allows to describe, at least qualitatively, the phase diagram
of doped CuGeO$_3$.

We are grateful to J.~Knoester, A.~Lande, O.~Sushkov and
G.~Sawatzky for useful comments.  D.~Kh.  is especially grateful
to B.  B\"uchner for extremely useful discussions and for
informing him of many experimental results prior to publication.
This work was supported by the Dutch Foundation for Fundamental
Studies of Matter (FOM).

\newpage
\widetext
\section*{Figure Captions}

\noindent
FIG. 1. Simplified structure of CuO$_2$ chains in CuGeO$_3$.

\vspace{1cm}

\noindent
FIG. 2. Electronic orbitals of $Cu$, $O$, and $Ge$ relevant for
superexchange.

\vspace{1cm}

\noindent
FIG. 3. Schematic structure of CuO$_2$ - Ge skeleton in
CuGeO$_3$.

\vspace{1cm}

\noindent
FIG. 4. SP soliton in the strong coupling limit
(above) and in the continuum model (below).

\vspace{1cm}

\noindent
FIG. 5. (a) A string confining soliton to an impurity
(indicated by a triangle); (b) total potential acting on
the soliton near the impurity.

\vspace{1cm}

\noindent
FIG. 6. Dependence of the SP transition temperature on
the concentration of dopands for several values of interchain
coupling.

\vspace{1cm}

\noindent
FIG. 7. Antiferromagnetic correlations around the
soliton.


\begin{thebibliography}{99}
\bibitem{1}
E. Pytte, Phys. Rev. B {\bf 10}, 4637 (1974).

\bibitem{2}
L. N. Bulaevskii, A. I. Buzdin, and D. I. Khomskii, Solid
State Commun. {\bf 27}, 5 (1978).

\bibitem{3}
M. C. Cross and D. S. Fisher, Phys. Rev. B {\bf 19}, 402
(1979); M. C. Cross, Phys. Rev. B {\bf 20}, 4606 (1979).

\bibitem{4}
J. W. Bray {\em et al}, in {\em ``Extended Linear Chain
Compounds''}, ed. J.~S.~Miller, (Plenum, NY 1985), p.353.

\bibitem{5}
M. Hase, I. Terasaki, and K. Uchinokura, Phys. Rev. Lett.
{\bf 70}, 3651 (1993).

\bibitem{6}
J. E. Lorenzo {\em et al}, Phys. Rev. B {\bf 50}, 1278
(1994).

\bibitem{7}
H. Winkelmann {\em et al}, Phys. Rev. B {\bf 51}, 12884
(1995).

\bibitem{8}
B. B\"{u}chner {\em at al}, Phys. Rev. Lett. in press.

\bibitem{9}
M. Hase {\em et al}, Phys. Rev. Lett. {\bf 71}, 4059 (1993).

\bibitem{10}
J. P. Renard {\em et al}, Europhys. Lett. {\bf 30}, 475
(1995).

\bibitem{11}
Y. Sasago {\em et al}, preprint cond-mat/9603185.

\bibitem{12}
L. P. Regnault {\em et al}, Europhys. Lett. {\bf 32}, 579
(1995).

\bibitem{13}
W. Geertsma and D. Khomskii, Phys. Rev. B {\bf 54} (1996).

\bibitem{15}

G. Castilla, S. Chakravarty and V. J. Emery, Phys. Rev.
Lett. {\bf 75}, 1823 (1995).

\bibitem{16}
C. K. Majumdar and D. K. Ghosh, J. Math. Phys. {\bf 10},
1388, 1899 (1969).

\bibitem{17}
W. Geertsma and D. Khomskii, to be published.

\bibitem{18}
M. Braden {\em et al}, preprint 1996.

\bibitem{19}
B. B\"uchner, W. Geertsma and D. Khomskii, to be published.

\bibitem{20}
K. Hirota {\em et al}, Phys. Rev. Lett. {\bf 31}, 736
(1994).

\bibitem{21}
A. J. Heeger, S. Kivelson, J. R. Schrieffer, and W.~P.~Su, Rev.
Mod. Phys. {\bf 60}, 781 (1988).

\bibitem{SU}
W.~P.~Su, Solid State Commun. {\bf 35}, 899 (1980).

\bibitem{22}
M. Mostovoy and D. Khomskii, to be published.

\bibitem{MFK}
M. Mostovoy, M. T. Figge, and J. Knoester, to be published.

\bibitem{23}
The coexistence of the SP and the antiferromagnetic phases was
recently treated by  H. Fukuyama, T. Tanimoto and M.
Saito, J. Phys. Soc. Japan, {\bf 65}, 1182 (1996). Our approach is
rather similar, with one important difference: due to the presence
of solitons the change of phase by $\pi$ is allowed in our
picture, whereas it is not realized in the solution obtained in
the above cited paper.
\end{thebibliography}
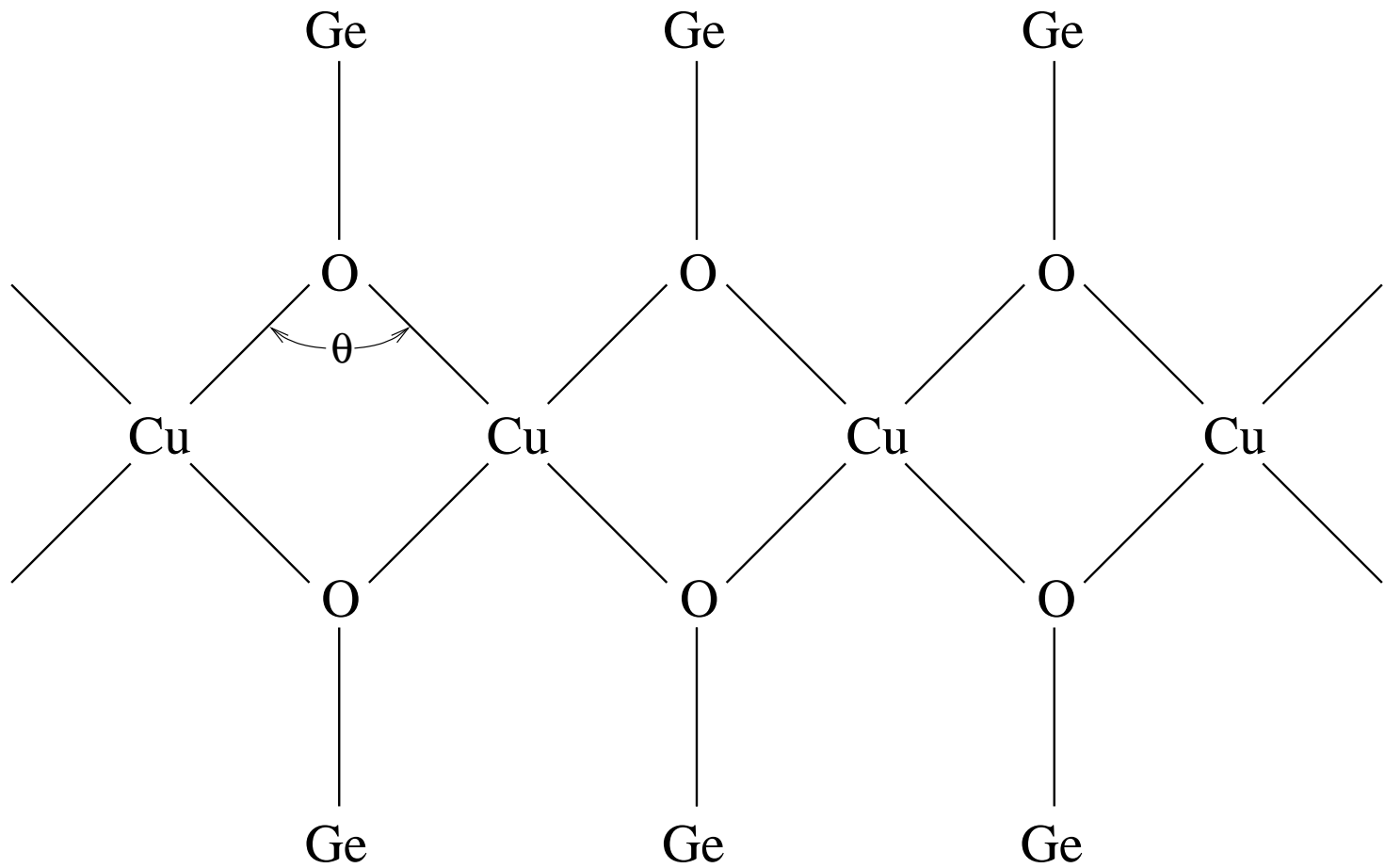
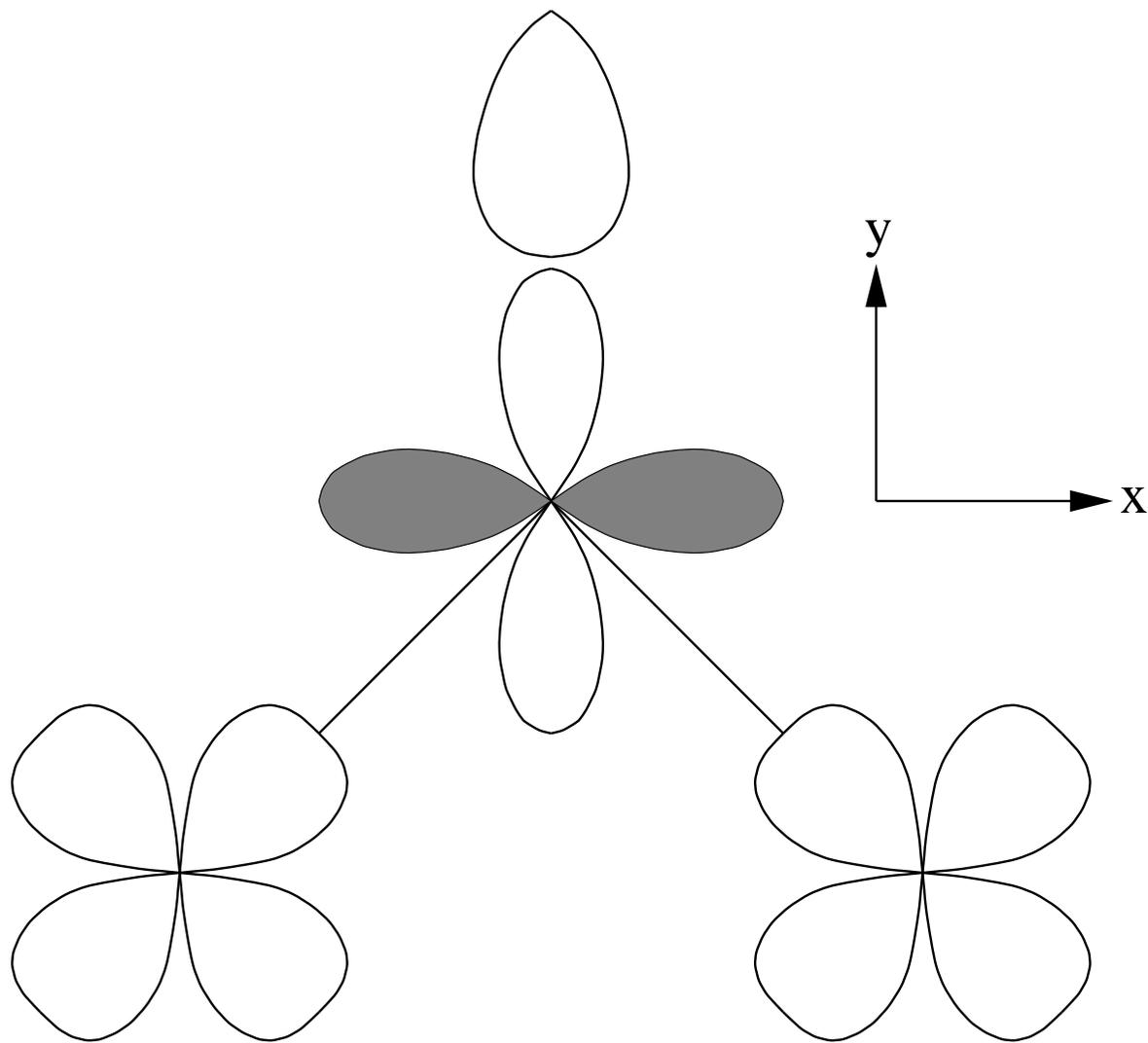
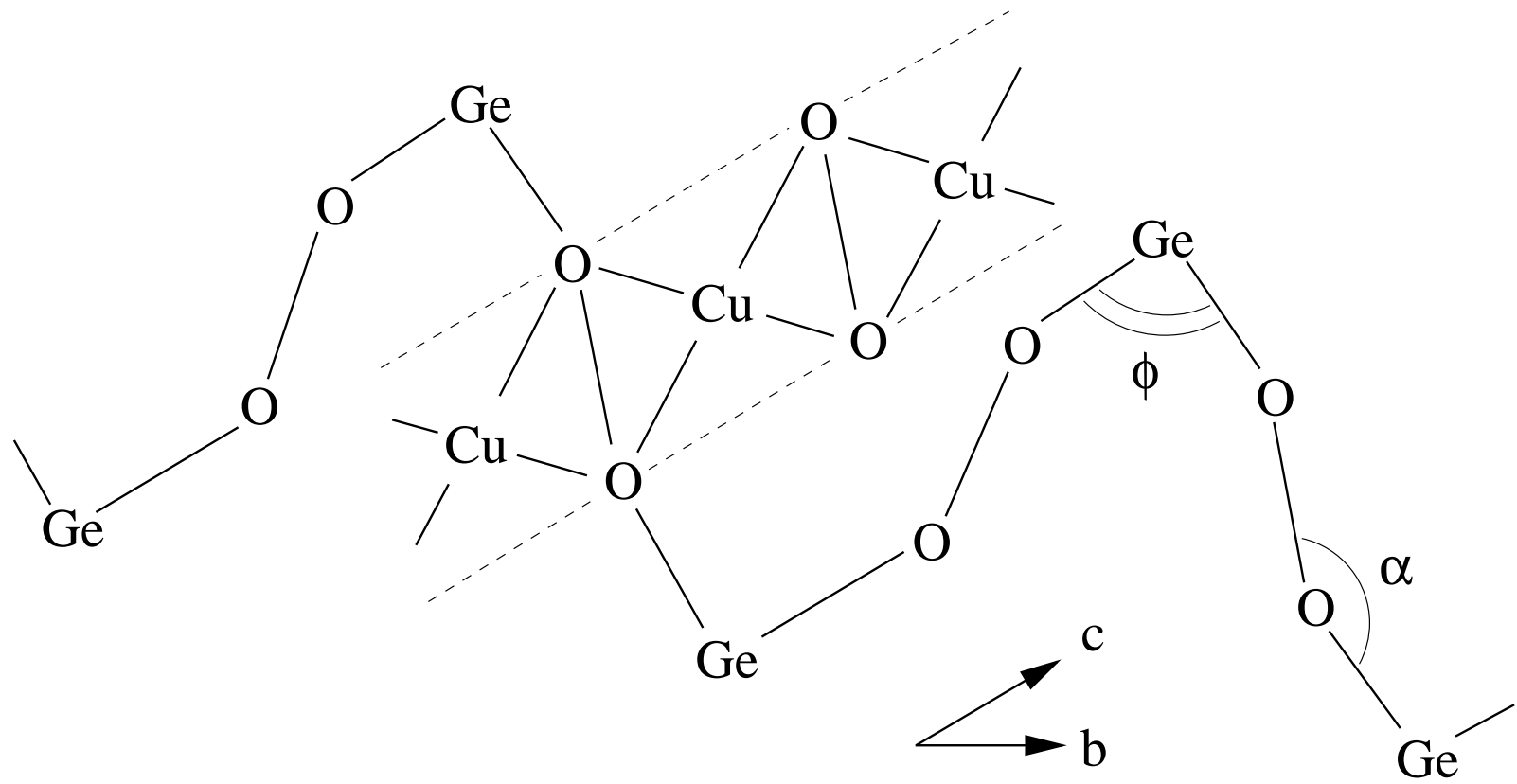
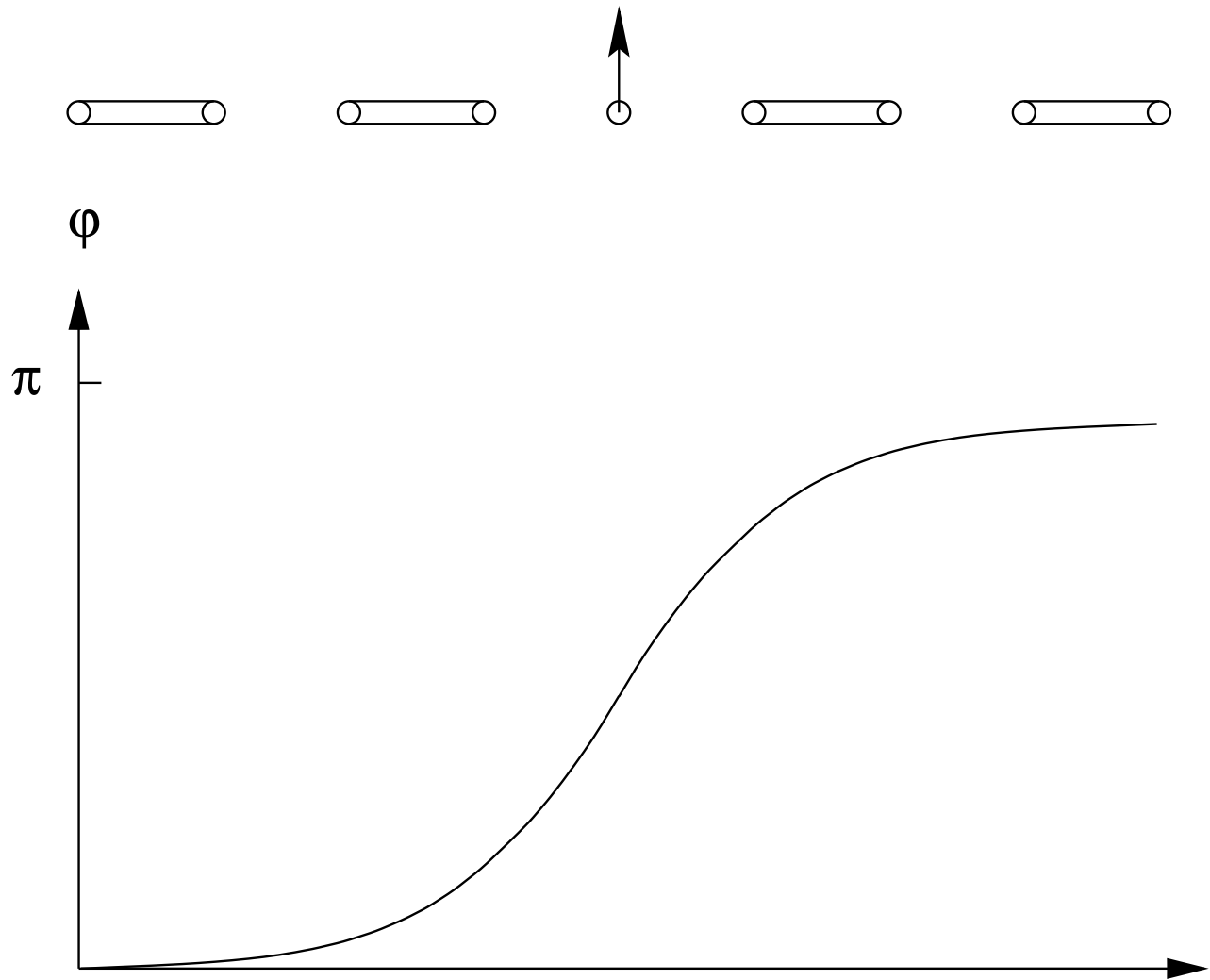
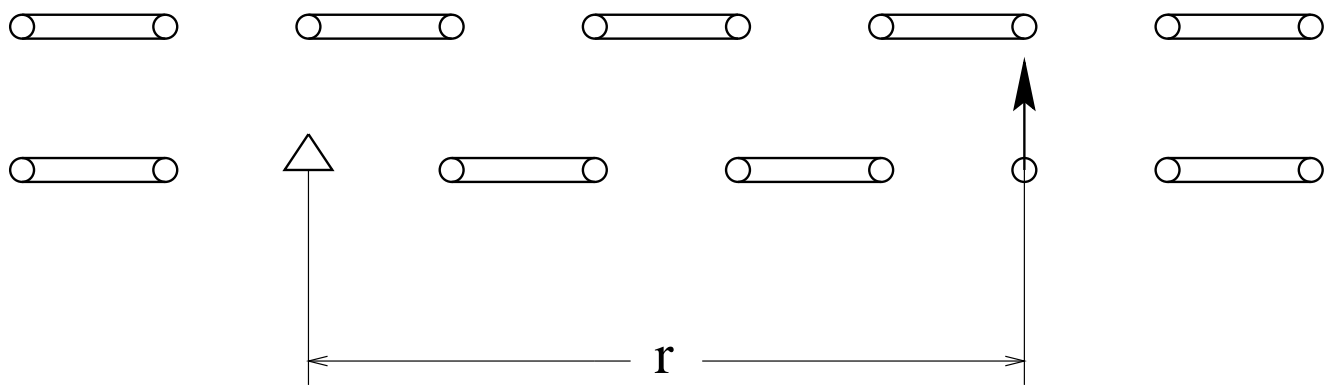
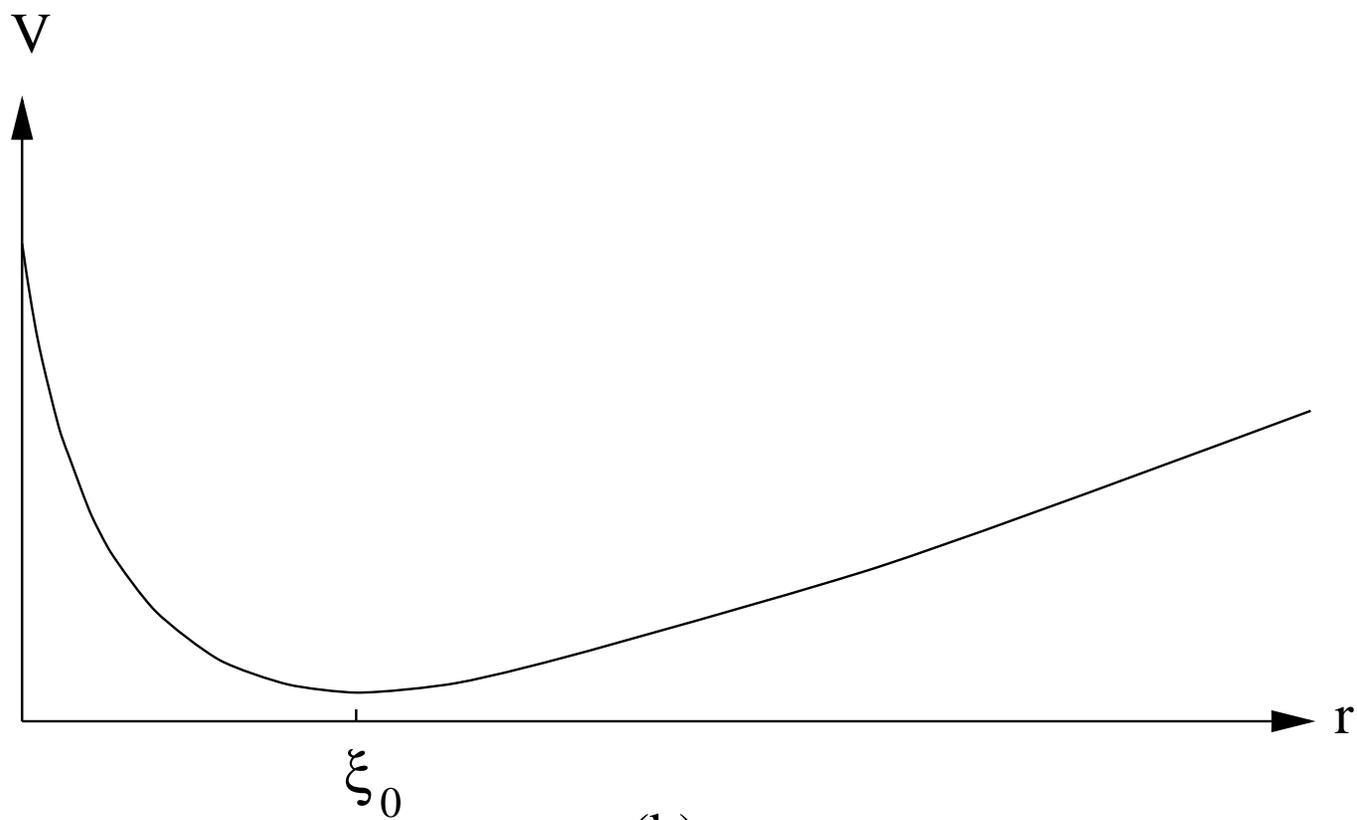
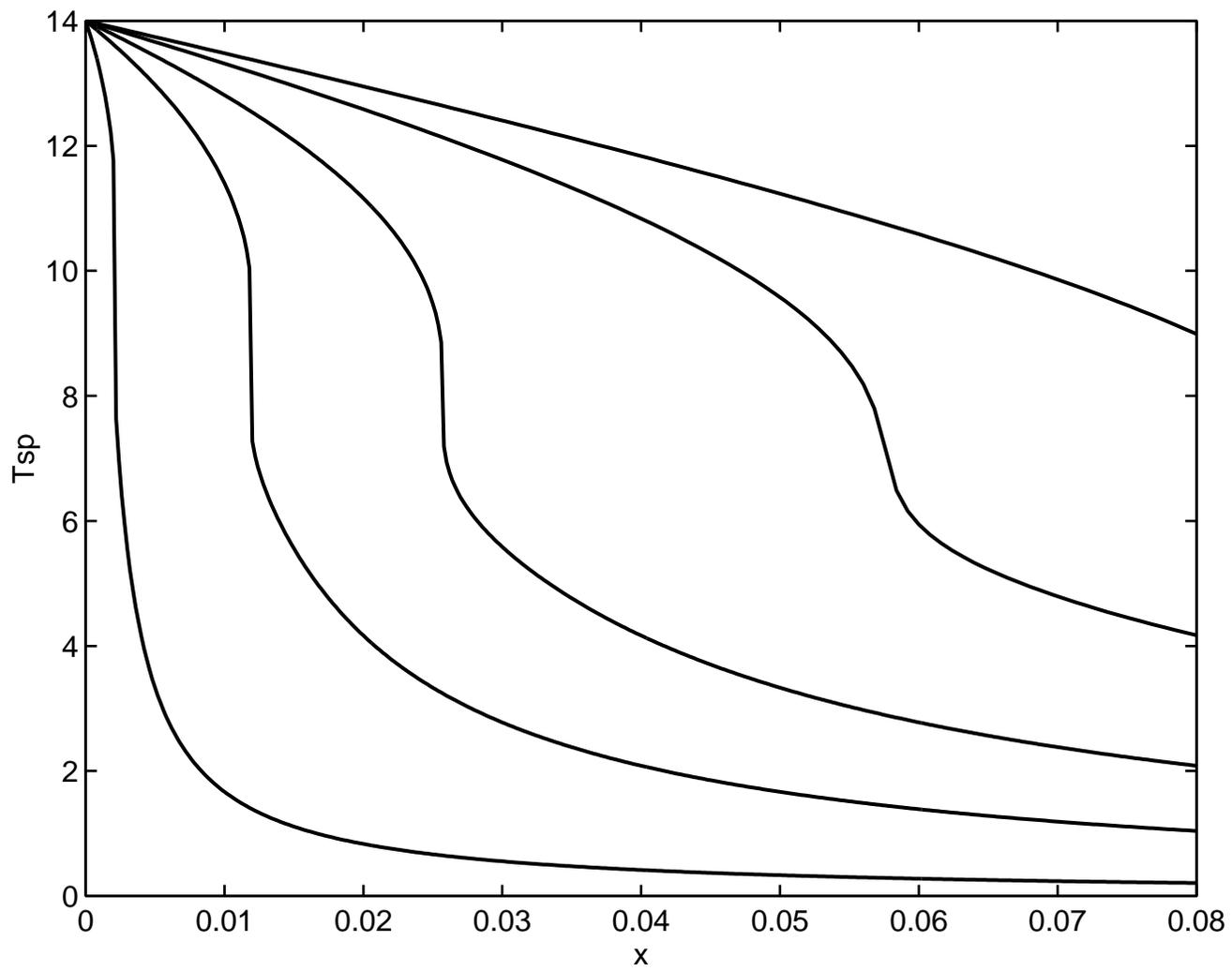
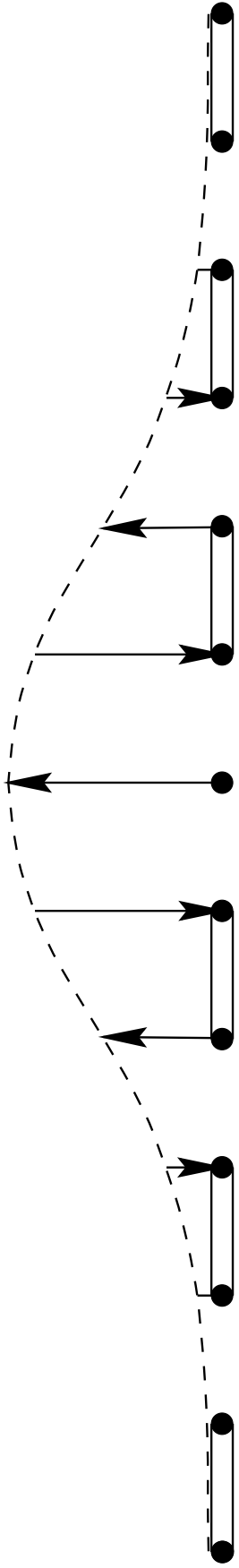
\end{document}